\input epsf.tex
\epsfclipon
 \documentclass[11pt,a4paper,amsart]{article}
 \usepackage{jheppub}

\usepackage{hyperref}

\usepackage{mathtools}
\usepackage{color}

\usepackage{epsfig}
\usepackage{epstopdf}
\usepackage{epsf}
\input{epsf.sty}
\usepackage{graphicx,amsmath,amssymb}
\usepackage{graphicx}
\usepackage{epsfig,multicol}
\usepackage{multirow}
\allowdisplaybreaks

\usepackage{tikz} 

\usepackage{bbm,bm,amsmath,amssymb}
 
\newcommand{\Su}{\color{blue}}

\usepackage[normalem]{ulem}

\usepackage{comment}
\def\be{\begin{equation}}
\def\ee{\end{equation}}
\def\figs/B{B}
\def\bea{\begin{eqnarray}}
\def\eea{\end{eqnarray}}
\def\bg{\begin{eqnarray}}
\def\nd{\end{eqnarray}}
\def\sin{{\rm sin}}
\def\cos{{\rm cos}}

\def\be{\begin{equation}}
\def\ee{\end{equation}}

\def\doi{http://doi.org}

\usepackage{physics}
\usepackage{tikz}
\usetikzlibrary{arrows.meta}
\usetikzlibrary{angles,quotes} 
\usetikzlibrary{decorations.markings,arrows.meta}
\tikzset{>=latex} 

\tikzset{
  midarr/.style={decoration={markings,mark=at position #1 with {\arrow{stealth}}},postaction={decorate}},
  midarr/.default=0.5
}
\colorlet{xcol}{blue!70!black}

\title{de Sitter State in Heterotic String Theory}
\author{{\Su Stephon Alexander}$^{1}$, {\Su Keshav Dasgupta}$^{2}$, {\Su Archana Maji}$^3$, {\Su P. Ramadevi}$^3$ and
{\Su Radu Tatar}$^{4}$\\
	\vskip.03in
        ${}^1$ Brown Theoretical Physics Center and Department of Physics, Brown University, Providence, RI 02912, USA\\ 
	${}^2$ Department of Physics, McGill University, Montr\'{e}al, Qu\'{e}bec, H3A 2T8, Canada \\
	${}^3$ Department of Physics, Indian Institute of Technology Bombay, Mumbai 400076, India\\
       ${}^4$ Department of Mathematical Sciences,
        University of Liverpool,  Liverpool, L69 7ZL, United Kingdom \\	{\tt stephon${}_-$alexander@brown.edu, keshav@hep.physics.mcgill.ca} 
        {\tt archana${}_-$phy@iitb.ac.in, ramadevi@iitb.ac.in, Radu.Tatar@Liverpool.ac.uk}}

\date{\today}

\abstract{Recent no-go theorems have ruled out four-dimensional classical de Sitter vacua in heterotic string theory. On the other hand, the absence of a well-defined Wilsonian effective action and other related phenomena also appear to rule out such time-dependent vacua with de Sitter isometries, even in the presence of quantum corrections. In this note, we argue that a four-dimensional de Sitter space can still exist in $SO(32)$ heterotic string theory as a Glauber-Sudarshan state, {\it i.e.} as an excited state, over a supersymmetric Minkowski background, albeit within a finite temporal domain. Borel resummation and resurgence play a crucial role in constructing such a state in the Hilbert space of heterotic theory governed entirely by the IR degrees of freedom.} 

\begin{document}

\maketitle

\section{Introduction and summary}	
\label{sec:intro}

Our modern understanding of quantum field theories is based on two recurring themes, one, on the existence of a Wilsonian effective action and two, on the asymptotic nature of the perturbation series. The latter, which was actually known for some time now \cite{dyson}, was surprisingly only appreciated more recently from some remarkable works \cite{unsal} which showed clearly how non-perturbative effects manifest themselves {\it naturally} in correlation functions. 

Extending both these themes to cosmological set-up wherein temporal dependences appear automatically is much more non-trivial. Even more challenging is the scenario where string theory is involved. In string theory, where the asymptotic nature of string perturbation theory is well documented, the existence of a Wilsonian effective action over a temporally varying cosmological background is not guaranteed. In fact there are strong evidences to suggest that a Wilsonian effective action may not exist because of the temporal dependences of the fluctuating frequencies, as well as of the massive stringy and the KK modes. For such a background, although we do expect {\it some} (as yet unknown) stringy description, it is a futile affair to search for a {\it supergravity} description where none exists. Equally futile then is the search for a vacuum solution for a cosmological background. These and other related arguments form the core of the so-called trans-Planckian problems in string-cosmology \cite{tcc}.

The situation, unfortunate as may seem, is not without hope. Solutions do exist, but not in a way envisioned earlier. Demanding the existence of a Wilsonian effective action then instructs us to realize the cosmological background $-$ which is a de Sitter space in this case $-$ as an excited state over a supersymmetric Minkowski background in string theory. We expect the excited state to break supersymmetry spontaneously, but the question is what kind of excited state are we looking for?  Clearly, since the universe we live in is very close to a {\it classical} one, the only excited state that has any chance of reproducing the classical behavior is a coherent state which amounts to shifting the free vacua in field theory. Unfortunately, due to the non-existence of free vacua in string theory, such a state cannot be easily realized and the closest we can come to realizing a coherent state would be by shifting the interacting vacua. This actually turned out to be a reasonably viable option as amply demonstrated for the type IIB case in \cite{coherbeta, coherbeta2, borel2, joydeep}, and we called such a state as the {\it Glauber-Sudarshan state} to distinguish it from the coherent state\footnote{Although we shall use both the terminologies variably throughout, it is the former that is always meant.}. 

Our aim in this paper is to realize a de Sitter state in heterotic string theory. Due to various technical reasons, $SO(32)$ heterotic theory appears to provide a more controlled laboratory than $E_8 \times E_8$ theory to implement the computational technology. This computational technology involves performing a full-fledged path-integral along the lines of \cite{coherbeta, coherbeta2, borel2, joydeep, borel3} over a Minkowski minimum which, expectedly, leads us to an asymptotic series of the Gevrey kind \cite{gevrey} thus requiring Borel resummation \cite{borelborel}. The final answer we get matches somewhat with the type IIB case from \cite{coherbeta, coherbeta2, borel2, joydeep}, but the details are quite different. These differences are important and in section \ref{sec2.3} we will spell them out. In \cite{hete8} we will elaborate further on the $SO(32)$ as well as the $E_8 \times E_8$ theories.

The note is organized in the following way. In section \ref{sec2.1} we point out the reason for uplifting the type IIB or the dual heterotic background to M-theory. In section \ref{sec2.2} we provide the duality chain that relates a type IIB orientifold background to a heterotic $SO(32)$ background, and in section \ref{sec2.3} we present our main results of constructing the de Sitter Glauber-Sudarshan state using Borel resummation of a Gevrey series and point out the key differences from the type IIB case. We end with a discussion in section \ref{sec3}.

\section{Quantum corrections, Glauber-Sudarshan state and M-theory \label{sec2}}

In \cite{coherbeta, coherbeta2} we showed how a four-dimensional type IIB background with de Sitter isometries can be realized as a Glauber-Sudarshan state. (For a non-technical introduction to the Glauber-Sudarshan states, the readers may look up section 8.1 in \cite{joydeep}.) As we also discussed therein, this background {\it cannot} appear as a {\it vacuum} configuration in IIB string theory due to numerous issues. The question that 
we want to ask here is that whether a generic background of the form:

{\footnotesize
\bg\label{viomyer1}
ds^2 = {a^2(t)\over {\rm H}^2(y)}\left[-dt^2 + g_{ij}dx^i dx^j + g_{33} (dx^3)^2\right] + 
{\rm H}^2(y)\left[{\rm F}_2(t) g_{mn} dy^m dy^n + {\rm F}_1(t) g_{\alpha\beta} dy^\alpha dy^\beta\right], 
\nd}
\noindent can also be realized as a Glauber-Sudarshan state. Here ${\rm F}_i(t)$ captures the dominant temporal scalings, and in what sense they do will be elaborated when we lift this configuration to M-theory. 
Note that $a^2(t)$, with $t$ being the dimensionless conformal time (measured with respect to ${\rm M}_p = 1$), is kept arbitrary with the only condition being that it becomes large at late time. This means the background \eqref{viomyer1} naturally {\it expands} at late time. For example when $a^2(t) = {1\over \Lambda t^2}$, with $\Lambda$ being the cosmological constant, we get an expanding de Sitter space in the flat slicing in type IIB as $t \to 0$ at late time.
The other factors, $g_{ij}({\bf x}), g_{33}({\bf x}), g_{mn}(y), g_{\alpha\beta}(y)$ and ${\rm H}^2(y)$ are the unwarped spatial metric components and the warp-factor respectively. The coordinate $y \equiv (y^m, y^\alpha) \in {\cal M}_4 \times {\cal M}_2$ and $x = (t, {\bf x}) \in {\bf R}^{2, 1}$, so that nothing depends on the third spatial direction parametrized by $x^3$ here. We will soon make a further restriction by converting $g_{\alpha\beta} = \delta_{\alpha\beta}$, so that ${\cal M}_2 = {\mathbb{T}^2\over \mathbb{Z}_2}$ where $\mathbb{Z}_2 \equiv \Omega (-1)^{\rm F_L} {\bf I}_{\mathbb{T}^2}$ will be an orientifolding 
operation (${\bf I}_{\mathbb{T}^2}$ is the orbifold action. Details are in \cite{DRS}). Such a choice will give us a way to reach the heterotic background by making a series of duality transformations\footnote{The duality transformations, as we will discuss soon, involve both T and S-dualities so one might be worried that such transformations are ill-defined over a non-supersymmetric background. We would like to assure the readers that this is {\it not} the case because the backgrounds (in both IIB and heterotic theories) are realized as Glauber-Sudarshan states over {\it supersymmetric} Minkowski spaces. One may then address the duality transformations in two ways \cite{hete8}. In the first way, we make duality transformations over the supersymmetric backgrounds and realize the corresponding non-supersymmetric spaces as Glauber-Sudarshan states over the duality transformed backgrounds and satisfying the Schwinger-Dyson equations \cite{joydeep}. In the second way, we could make the duality transformations over the non-supersymmetric backgrounds but keep track of all-order $\alpha'$ and $g_s$ corrections. Both lead to the same answer and are surprisingly tractable as will be shown in our upcoming paper \cite{hete8}. For the present work, we will avoid these subtle nuances. \label{sosie}}. In that case $y \equiv y^m \in {\cal M}_4$. For the time being, however, we will continue with the generic picture. 

The reason for this genericity is simple. As alluded to above, for various choices of $\left(a^2(t), {\rm F}_i(t)\right)$ and the internal sub-manifolds, we can study the possibilities of realizing de Sitter state in various string theories (including also in M-theory). Such realizations will involve duality transformations: for example appropriate T-dualities, with and without any orientifolding operations, can give rise to the possibiltites of de Sitter states in type I and type IIA theories respectively. With an additional S-duality, as mentioned earlier, we could study de Sitter state in heterotic $SO(32)$ theory (appropriately broken to a suitable subgroup). 
We can even dualize to M-theory (see \cite{coherbeta2}) and from there dualize further to heterotic $E_8 \times E_8$ theory. The questions that we want to investigate here is whether such possibilities could be explicitly realized. 

Expectedly, there are also a few other changes from the construction in \cite{coherbeta, coherbeta2}. We no longer impose any constraint on ${\rm F}_i(t)$. This means the four-dimensional Newton constant {\it can} become time-dependent in the IIB side. What we do want, however, is that the Newton constant remains time-{\it independent} in the dual side ({\it i.e.} the dual side where we want to realize the de Sitter state). In a similar vein, the functional form for $a^2(t)$ will be determined by demanding a de Sitter space in the dual side. The precise conditions on $a^2(t)$ and ${\rm F}_i(t)$ will be elucidated once we dualize to the corresponding theory. 

The dualities to the various string and M-theory sides are more subtle now because the type IIB background \eqref{viomyer1} cannot be realized {\it classically} \cite{GMN}. Quantum mechanically we expect such a background to exist only in the presence of all possible perturbative, non-perturbative and non-local, including topological corrections \cite{joydeep}. Additionally $-$ on one hand $-$ temporal dependences of the underlying degrees of freedom are absolutely essential for an Effective Field Theory (EFT) to exist. (The existence of EFT is in turn related to the existence of four-dimensional Null Energy Condition (NEC) \cite{coherbeta2}.) On the other hand, existence of the temporal degrees of freedom, for example fluxes etc., are tightly constrained by the flux quantization and anomaly cancellation conditions. Thus the system is highly intertwined, and unless we demonstrate that a background like 
\eqref{viomyer1} can exist (at least as a Glauber-Sudarshan state), the duality chasing will be a meaningless
exercise.  

The last comment on the existence of the background \eqref{viomyer1} as a Glauber-Sudarshan state deserves some explanation. As we saw in \cite{coherbeta, coherbeta2}, when $a^2(t)$ specifies a given de Sitter slicing, Wilsonian effective action can only be defined properly when the background becomes a Glauber-Sudarshan state\footnote{As alluded to in the introduction, the key differences between the Glauber-Sudarshan state and the standard coherent state are described  in \cite{coherbeta, coherbeta2, joydeep}. We will not go into those details here and the readers may pick up all the relevant informations from the references, and especially from \cite{joydeep}.}. Other issues like the existence of a Trans-Planckian Cosmic Censorship (TCC), moduli stabilization, Faddeev-Popov ghosts, Schwinger-Dyson equations etc., appear much more naturally in this framework. More so, the existence of the Glauber-Sudarshan state tells us how a de Sitter state may exist in the type IIB string {\it landscape} (and not in the so-called {\it swampland}).  The question that we want to ask here is whether such a de Sitter Glauber-Sudarshan state can be found in the dual landscape. The answer, as we shall see, turns out to be more complex
in a sense that shall be elaborated soon\footnote{For example, one of the question whose answer that we seek is as follows. Since heterotic and type IIB are related by a set of duality transformations, doesn't that naturally guarantee a de Sitter state in the heterotic side? The answer is {\it no}. In fact, as we shall see, duality chasing only works if the seed background in the IIB side exists. Our analysis will hopefully reveal that this is {\it not} guaranteed {\it a-priori}.}. But first: does the background \eqref{viomyer1} exist in the IIB landscape as a Glauber-Sudarshan state? This is what we turn to next.

\subsection{Consistency of M-theory uplift and Glauber-Sudarshan state \label{sec2.1}}

As in \cite{coherbeta, coherbeta2, joydeep}, we will lift the background \eqref{viomyer1} to M-theory by T-dualizing along $x^3$ and then uplifting the configuration to eleven-dimensions. There are various reasons why such an uplift becomes {\it necessary}. 

\vskip.1in

\noindent $\bullet$ The type IIB background \eqref{viomyer1} is supported at a constant coupling point in F-theory \cite{senkd}. This means axio-dilaton vanishes and the IIB coupling $g_b = 1$. This is a strong coupling point where S-duality doesn't help. More so, the vanishing axio-dilaton for example is necessary to get a gauge group of 
${\bf D}_4^4 = \left[{\rm SO}(8)\right]^4$ in the dual heterotic side. The $E_8$ heterotic theory appears in a more non-trivial way as shown in \cite{horava}.

\vskip.1in

\noindent $\bullet$ The internal space with topology ${\cal M}_4 \times {\cal M}_2$ is not only a non-K\"ahler manifold but is also non-complex. Additionally, various parts of the internal space evolve differently with time, as shown by the temporal factors ${\rm F}_i(t)$. The space-time has positive cosmological constant, and is therefore highly non-supersymmetric\footnote{If we can realize the background as a Glauber-Sudarshan state over a supersymmetric Minkowski space (with a non-K\"ahler non-complex internal manifold), much like in \cite{desitter2, coherbeta, coherbeta2, joydeep}, then the supersymmetry is broken spontaneously. The supersymmetric vacuum appears from the self-dual G-fluxes. Once we take the expectation values of the G-fluxes over the Glauber-Sudarshan state, they no longer remain self-dual and therefore break supersymmetry spontaneously. See \cite{coherbeta, coherbeta2, joydeep} for details.}.  Putting everything together we see that all conventional techniques that we have learnt so far would fail to quantify the dynamics of the system.

\vskip.1in

\noindent $\bullet$ The unit coupling in the IIB side means that no controlled computation can be performed there. The only leverage we can get is by dualizing to the IIA side where the IIA coupling $g_s$ becomes 
${g_s\over {\rm HH}_o} = {1\over a(t)}$, where ${\rm H}(y)$ is the warp-factor and ${\rm H}_o({\bf x})$ is related to $g_{33}({\bf x})$. For an expanding cosmology, the system then is naturally {\it weakly coupled} at late times. 

\vskip.1in

\noindent $\bullet$ Uplifting this to M-theory converts most of the IIB fluxes to four-form G-fluxes. Moreover, the IIB seven-branes become geometric spaces in M-theory \cite{dileep}. The D3-branes, which are instantons on the seven-branes, naturally also become geometric. The temporal dependences of the G-flux components will make the D3-branes {\it dynamical}. 

\vskip.1in

\noindent $\bullet$ In the small-instanton limit, these D3-branes dualize to either D2 or D4-branes in IIA, that are uplifted to M2 and M5-branes respectively. Since both D2 and D4-branes dissolve as instantons or as first Chern classes respectively on the IIA D6-branes, they naturally become {\it dynamical} in M-theory\footnote{Recall that the gauge fluxes on the D6-branes appear from localized G-fluxes of the form ${\bf G}_{{\rm MN}ab}$ and ${\bf G}_{{\rm 0M}ab}$ that are generically time-dependent \cite{desitter2, coherbeta, coherbeta2} (the notations are explained after \eqref{amhomes}). This would make both the world-volume gauge fluxes as well as the branes dynamical.}. These dynamical M5-branes are responsible for the flux quantization procedure as shown in \cite{coherbeta2, desitter2}.

\vskip.1in

\noindent $\bullet$ The IIB D-string dualize to either D0 or D2-branes in IIA. The D0-branes uplift to the massive gravity multiplet (which makes sense because the magnetic dual of the D0-branes, namely the D6-branes, dualize to Taub-NUT spaces (or KK-monopoles) in M-theory). The type IIB fundamental string dualize to a wrapped M2-brane when uplifted. 

\vskip.1in

\noindent $\bullet$ The Glauber-Sudarshan states are most succinctly presented from M-theory perspective as they could easily reproduce any metric and flux configurations. Since most of the IIB branes dualize to either geometry or flux configurations when uplifted to M-theory, the Glauber-Sudarshan states could in principle reproduce these configurations too. From the IIB side we could probably view them from a string field theory set-up, where the shifting of the interacting IIB vacuum could be naturally realized\footnote{This shifting is of course an essential ingredient in constructing the Glauber-Sudarshan states. In M-theory this is
easily realized at low energies and is directly related to the wave-function renormalization of the usual 
coherent states. There are quite a few subtleties that we have kept under the rug here which the readers could get from section 6.1 of the first reference in \cite{coherbeta2}.}. 

\vskip.1in

\noindent $\bullet$ Non-existence of a well-defined action in the IIB side is also another reason for the uplift to M-theory. As we saw in \cite{desitter2, coherbeta, coherbeta2, joydeep} it is absolutely essential to spell out the precise set of perturbative, non-perturbative, non-local and topological quantum corrections. It is {\it only} in M-theory that such a procedure may be explicitly performed. Existence of a Wilsonian effective action $-$ as guaranteed from the existence of a Glauber-Sudarshan state\footnote{As demonstrated carefully in \cite{coherbeta, joydeep}, these two facts go hand-in-hand. One implies the other. This is also a stronger reason for viewing de Sitter space as a Glauber-Sudarshan state and not as a vacuum configuration.} $-$ means that all the short-distance degrees of freedom can be integrated out to express the quantum corrections in the form given in \cite{desitter2, coherbeta, coherbeta2, joydeep}. 

\vskip.1in

\noindent $\bullet$ Lifting the metric configuration \eqref{viomyer1} to M-theory, one may easily see that the toroidal direction, parametrized by $(x^3, x^{11})$, scales as $\left({g_s\over {\rm HH}_o}\right)^{4/3}$ which becomes arbitrarily {\it small} in the limit $g_s \to 0$. Since we are always in the limit\footnote{The reason for this is to control the non-perturbative corrections of the form ${\rm exp}\left[-\left({g_s\over {\rm HH}_o}\right)^{-2k/3}\right]$, where $k \in \mathbb{Z}_+$. So while M-theory is usually defined for $g_s >> 1$, we want to be in the opposite limit \cite{joydeep}.} $g_s < 1$, the M-theory degrees of freedom do capture the type IIB behavior exactly. This also means that, for example if we are allowed to keep the ${\cal M}_2$ cycle small compared to $\alpha'$ we can, in the orientifold limit, capture the $SO(32)$ heterotic dynamics using the M-theory degrees of freedom. Thus the M-theory uplift has a dual advantage: provide both the IIB and the heterotic dynamics under appropriate conditions. In fact, as it turns out, the M-theory configuration is a {\it master} theory from where the de Sitter dynamics can be determined for all the string theories (including de Sitter state directly in M-theory). The proof of the statement is beyond the scope of this paper and will be demonstrated elsewhere.

\vskip.1in

\noindent Interestingly, the {\it form} of the M-theory metric always remains the same for any type IIB cosmology expressed using conformal coordinates as in \eqref{viomyer1}. The only change is the value of the dual IIA string coupling: ${g_s\over {\rm HH}_o} = {1\over a(t)}$ which is sensitive to the functional form of $a(t)$. Clearly, and as mentioned earlier, for expanding cosmologies the IIA coupling can be made small. In fact demanding ${g_s\over {\rm HH}_o} < 1$, provides the temporal domain in which controlled quantum computations may be performed in M-theory. For the usual de Sitter case, irrespective of the choice of the de Sitter slicings, this temporal domain remains perfectly consistent with the so-called Trans-Planckian Cosmic Censorship (TCC) \cite{tcc}, as shown in \cite{desitter2, coherbeta, coherbeta2, joydeep}. We now proceed to find the functional form of $a(t)$ that provides a de Sitter metric in the {\it dual} heterotic side.  In what follows we elaborate on this solution.

\subsection{M-theory uplift of a heterotic $SO(32)$ background and dualities \label{sec2.2}}

The duality from type IIB to heterotic $SO(32)$ theory, in the presence of orientifolds and fluxes, has been explicitly shown in
\cite{DRS}. We will basically follow similar duality chasing here too, 
but not before we elucidate the consistency of the IIB background \eqref{viomyer1} from M-theory.
In M-theory, the uplifted metric takes the following standard form:

{\footnotesize   
\bg\label{amhomes}
ds^2 &= & \left({g_s\over {\rm H}{\rm H}_o}\right)^{-8/3}\left(-\tilde{g}_{00} dt^2 + \tilde{g}_{ij} dx^i dx^j\right) 
+ \left({g_s\over {\rm H}{\rm H}_o}\right)^{-2/3} 
\bigg[{\rm F}_1\left({g_s\over {\rm H}_1}\right) \tilde{g}_{\alpha\beta} dy^\alpha dy^\beta 
+ {\rm F}_2\left({g_s\over {\rm H}_1}\right) \tilde{g}_{mn} dy^m dy^n\bigg]\nonumber\\
&&~~~~~~~ + \left({g_s\over {\rm H}{\rm H}_o}\right)^{4/3} \tilde{g}_{ab} dw^a dw^b,\nd}
where ${\rm H}_1({\bf x}, y) \equiv {\rm H}(y){\rm H}_o({\bf x})$, which means ${\rm F}_i(g_s/{\rm H}_1)$ depends on the temporal factor $a(t)$, and we shall discuss their functional form soon. (See footnote \ref{sosie} for a discussion on the duality transformation.) The other metric components may be related to the metric components in \eqref{viomyer1} in the following way:
\bg\label{morgcome}
&& \tilde{g}_{ab}({\bf x}, y) \equiv \Big[{\rm H}(y){\rm H}_o({\bf x})\Big]^{4/3}g_{ab}({\bf x}, y)\nonumber\\
&&\tilde{g}_{\mu\nu}({\bf x}, y) \equiv {g_{\mu\nu}({\bf x}) \over \Big[{\rm H}^4(y){\rm H}_o({\bf x})\Big]^{2/3}}, ~~~~
\tilde{g}_{\rm MN}({\bf x}, y) \equiv \left[{{\rm H}^2(y) \over {\rm H}_o({\bf x})}\right]^{2/3} g_{\rm MN}(y)\nd
where we have taken ${\rm M, N} \in {\cal M}_4 \times {\cal M}_2$ and 
$(w^a, w^b) \equiv (x^3, x^{11})$. Note that we have taken the un-warped metric components along the toroidal direction to depend on both $(x^i, y^{\rm M})$. In fact, for the computations of the curvature scalings, as shown in \cite{desitter2, coherbeta2}, one may take both the un-warped and the warped metric components to depend on all the coordinates (except of course the toroidal direction). Once we go to the heterotic side, we will see that the dependence on the coordinates of ${\cal M}_2$ have to be removed. 

Let us now come to the functional form for the temporal factors ${\rm F}_i(g_s/{\rm H}_1)$. In our earlier works \cite{desitter2, coherbeta, coherbeta2}, these factors did not change the dominant scalings of the metric components as they were constrained by ${\rm F}_i(g_s/{\rm H}_1) \to 1, g_s \to 0$ and ${\rm F}_1 {\rm F}_2^2 = 1$ to preserve the Newton's constant and to avoid late-time singularities. Both of these conditions are not essential now if we want to dualize to any of the other string and M-theories because only in the {\it dual} landscape we want a time-independent Newton's constant with no late time singularities.  This means the dominant scalings of the internal metric could in-principle {\it change}, implying changes to the curvature scalings from what we had in \cite{desitter2, coherbeta, coherbeta2}. We can then propose the following scalings:

{\footnotesize
\bg\label{sezoe70}
{\rm F}_1 \equiv \sum_{k = 0}^\infty {\rm A}_k\left({g_s\over {\rm HH}_o}\right)^{\beta_o + 2k/3}, ~~
{\rm F}_2 \equiv \sum_{k = 0}^\infty {\rm B}_k\left({g_s\over {\rm HH}_o}\right)^{\alpha_o + 2k/3}, ~~
{\partial\over \partial t}\left({g_s\over {\rm HH}_o}\right) \equiv \sum_{k = 0}^\infty {\rm C}_k\left({g_s\over {\rm HH}_o}\right)^{\gamma_o + 2k/3}, \nd}
where $({\rm A}_k, {\rm B}_k, {\rm C}_k)$ are all integers, positive or negative with $(\alpha_o, \beta_o, \gamma_o)$ being the dominant scalings
and $k \in \mathbb{Z}_+$. Note that, as we demonstrated rigorously in \cite{coherbeta2}, when $\gamma_o < 0$ EFT breaks down along-with a violation of the four-dimensional NEC. Here, in the generic setting, we will see whether this continues to hold or not. On the other hand $(\alpha_o, \beta_o)$ are not {\it a-priori} required to be positive definite. An interesting question would be to find whether there is a connection between the three dominant scalings. If there is one then it would lead to an even deeper connection between the three disparate facts: existence of EFT from M-theory, preserving four-dimensional NEC from IIB, and temporal dependence of the internal six-dimensional manifold. 

Before going into this, let us clarify couple of more things about the temporal dependence of the internal manifold. {\Su One}, taking \eqref{sezoe70} at face-value might imply that all components of the internal metric should scale in a certain way temporally. This is actually not the case {\rm as long as we maintain the dominant scalings}. For example we can start by generalizing the internal metric components as:
\bg\label{century}
&& {\bf g}_{\alpha\beta}({\bf x}, y; g_s) = \sum_{k = 0}^\infty {\rm A}_k^{(\alpha, \beta)}\left({g_s\over {\rm HH}_o}\right)^{-{2\over 3} + \beta_o + {2k\over 3}} {g}^{(k)}_{\alpha\beta}({\bf x}, y),\nonumber\\
&& {\bf g}_{mn}({\bf x}, y; g_s) = \sum_{k = 0}^\infty {\rm B}_k^{(m, n)}\left({g_s\over {\rm HH}_o}\right)^{-{2\over 3} + \alpha_o + {2k\over 3}} {g}^{(k)}_{mn}({\bf x}, y),  \nd
where the repeated indices are not summed over, and the $g_{\rm MN}^{(k)}({\bf x}, y)$ are the various possible metric components. It is easy to see that, unless we impose $g^{(k)}_{\rm MN}({\bf x}, y) \equiv 
\tilde{g}_{\rm MN}({\bf x}, y)$ from \eqref{morgcome}, it will in general be hard to keep the Newton's constant time-independent even in the type IIB side (although we are not required to do so now). Thus as long as the dominant scalings are divided into two sets: $(\alpha_o, \beta_o)$, the system can be generalized without violating the EFT constraints. Note that this implies that further splitting into three or more dominant scalings may {\it not} be necessary as higher order splittings can always be brought back to the two-splitting case by choosing appropriate values for 
${\rm A}_k^{(m, n)}$ and ${\rm B}_k^{(\alpha, \beta)}$. More details on this will appear in \cite{hete8}.

{\Su Two}, the {\it signs} of $(\alpha_o, \beta_o)$ are important. If $0 < \beta_o < +{2\over 3}$ and $\alpha_o \le 0$, it would mean that the size of the two-cycle ${\cal M}_2$ {\it shrinks} at late time when $g_s \to 0$ in the IIB side. Beyond certain short-distance scale, T-duality would become necessary, and if ${\cal M}_2 \equiv {\mathbb{T}^2\over \mathbb{Z}_2}$ $-$ with $\mathbb{Z}_2$ being the orientifold operation \cite{DRS} $-$ we see that the late time physics may be succinctly captured by either the Type I or the heterotic theory. On the other hand if $0 < \alpha_o < +{2\over 3}$ and $\beta_o \le 0$, the late time physics may still be given by the IIB theory, assuming no orientifolding operation, although problems like late-time singularities could arise in the absence of local isometries and in the presence of orientifolds. Note that we did not encounter any of these subtleties in \cite{coherbeta, coherbeta2} because both the space-time and the internal six-manifold had dominant scalings of $\left(-{8\over 3}, -{2\over 3}\right)$ and therefore the late-time physics was controlled by the low-energy effective action\footnote{This doesn't mean that the dynamics are controlled by classical supergravity as the late time physics is {\it not} necessarily captured by weakly curved manifolds, implying further that all perturbative, non-perturbative, non-local and topological corrections are {\it essential} to solve the EOMs as shown in \cite{desitter2, coherbeta, coherbeta2} and especially in \cite{joydeep}. The shrinking of the M-theory torus doesn't create any additional late-time singularities as we discussed earlier.}. 

With this we are almost ready to make the duality transformations to the heterotic side, keeping in mind the criteria of footnote \ref{sosie}. We will assume that $g_{\alpha\beta} = \delta_{\alpha\beta}$ specifies a square torus capturing
 the local metric of ${\cal M}_2 = {\mathbb{T}^2\over \mathbb{Z}_2}$, and $y \equiv y^m \in {\cal M}_4$ with 
${\cal M}_4$ being a generic non-K\"ahler manifold, as mentioned earlier. There are also NS and RR two-forms with components ${\bf B}_{\alpha m}(y, g_s)$ and 
${\bf C}_{\alpha m}(y, g_s)$ respectively with one of their legs along the toroidal directions. The heterotic metric then takes the following form\footnote{This metric was first derived by Evan McDonough in 2015 (and later by Bohdan Kulinich in 2022) by following the duality chasing argument mentioned above. We thank them for many discussions related to heterotic de Sitter solutions.}:
\bg\label{kimmetea}
ds^2_{\rm het} &= & {\rm F}_1(t) a^2(t) \left[-dt^2 + g_{ij}dx^i dx^j 
+  g_{33} (dx^3)^2\right] \\
&+ & {\rm H}^4(y){\rm F}_1(t) {\rm F}_2(t)~g_{mn} dy^m dy^n +
\delta_{\alpha\beta}\big(dy^\alpha + {\bf B}^\alpha_m dy^m\big)
\big(dy^\beta + {\bf B}^\beta_n dy^n\big), \nonumber \nd
where note that the toroidal directions no longer allow a warp-factor, but do allow a non-trivial fibration coming from the NS two-forms. Additionally, since the NS two-forms are time-dependent, the fibration naturally becomes time-dependent too. On the other hand, the temporal dependence of the RR two-forms provides the necessary {\it torsion} to support the non-K\"ahler, time-dependent metric \eqref{kimmetea} in the heterotic side. 

We now want the four-dimensional part of the metric \eqref{kimmetea} to be a de Sitter metric in some specific slicing. For simplicity we will choose a {\it flat slicing}. (We avoid {\it static patch} or any other slicings related to the static patch because of the issues mentioned in \cite{coherbeta2, joydeep}.) We also want the four-dimensional Newton's constant to be time-independent. Putting these two together implies that\footnote{Note that the volume ${\rm V}_6$ of the internal non-K\"ahler six-manifold in \eqref{kimmetea} is in general time-dependent but independent of the fibration structure and is given by ${\rm V}_6 = {\rm H}^8(y) {\rm F}^2_1(t) {\rm F}^2_2(t)
\sqrt{{\rm det}(g_{mn})}$. Once the second condition of \eqref{vitbeat} is taken into account, the volume becomes time-independent.}:
\bg\label{vitbeat}
a^2(t) = {1\over \Lambda t^2 {\rm F}_1(t)}, ~~~~~ {\rm F}_1(t) {\rm F}_2(t) = 1, \nd
in the original type IIB metric \eqref{viomyer1}, where $\Lambda$ is the cosmological constant, $t$ is the conformal time in flat-slicing and $(g_{ij}, g_{33}) = (\delta_{ij}, 1)$. (We can keep everything {\it dimensionless} by measuring with respect to ${\rm M}_p \equiv 1$ as shown in \cite{coherbeta, coherbeta2}.) The choice \eqref{vitbeat} tells us that the original type IIB metric \eqref{viomyer1} is now no longer required to have a time-independent four-dimensional Newton's constant, or have a metric with four-dimensional de Sitter isometries. The M-theory uplift however takes the same form as in \eqref{amhomes} but with the following choices for $g_s$, ${\rm H}_o({\bf x}), \alpha_o$ and $\beta_o$:
\bg\label{beatrvit}
{g_s\over {\rm HH}_o} = t \sqrt{\Lambda {\rm F}_1(t)}, ~~~~~ {\rm H}_o({\bf x}) = 1, ~~~~~ \beta_o = - \alpha_o, ~~~~~ 0 < \beta_o < {2\over 3}, \nd
showing that there is a dynamical possibility of the type IIB metric \eqref{viomyer1} to go to the heterotic side at late time. This also fixes the form for ${\rm F}_2(t)$. One might worry that this could in-principle over-constrain the system, but a similar computation in the type IIB side done in \cite{desitter2} shows that this may not be the case. Of course we will have to fix the functional form for ${\rm F}_1(t)$ using the Schwinger-Dyson type equations (in the presence of all non-perturbative and non-local corrections), but there is an additional condition on ${\rm F}_1(t)$ borne out of the non-violating-NEC criterion from \cite{coherbeta2}, namely:
\bg\label{voluntina}
{d{\rm F}_1\over dt} = {2\sqrt{{\rm F}_1}\over t \sqrt{\Lambda}}\left[\sum_{k = 0}^\infty 
{\rm C}_k \left(t \sqrt{\Lambda {\rm F}_1}\right)^{\gamma_o + 2k/3}  -  \sqrt{\Lambda{\rm F}_1}\right], \nd
where $\gamma_o$ is defined in \eqref{sezoe70}. Once we determine the form of ${\rm F}_1(t)$, the above equation will then fix the values of the constant coefficients ${\rm C}_k$ and $\gamma_o$. If $\gamma_o < 0$, then unfortunately such a system cannot be embedded in a UV complete theory, {\it i.e.} the solution (as a Glauber-Sudarshan state) cannot exist in heterotic string theory.

There are however reasons to believe that $\gamma_o$ could simply be zero if not a positive integer, but never negative. This is because the solution with $\beta_o = \alpha_o = 0$ has been studied earlier in \cite{desitter2}, where no apparent inconsistencies were detected. In the following section, we will demonstrate, under some mild approximations, that $\gamma_o$ can indeed be made positive definite here as long as $\beta_o \ge 0$.

\subsection{The resurgence of de Sitter space as a Glauber-Sudarshan state \label{sec2.3}}

Our aforementioned discussion should convince the readers that an M-theory uplift of the heterotic background \eqref{kimmetea} $-$ via type IIB $-$ is possible and is given by \eqref{amhomes} with the choice of parameters from \eqref{beatrvit}. Unfortunately however, such a background cannot be realised as a vacuum solution either classically or quantum mechanically. (By latter we mean the quantum corrections to the supergravity EOMs.) The former is ruled out in \cite{sethiK} from the constraints coming from the Kac-Moody algebras with $SO(4, 1)$ global symmetries\footnote{See also 
the last reference in \cite{GMN} for a different take on the no-go conditions stemming directly from the energy conditions.}. The latter is ruled out in \cite{coherbeta, coherbeta2} from the absence of a well-defined Wilsonian effective action for an accelerating background like \eqref{viomyer1} or \eqref{kimmetea}\footnote{One might argue that the situation could be dealt using {\it open} quantum field theories \cite{vernon} which is suited to tackle scenarios where energy is not conserved. However in string or M-theories it is not a priori clear how to separate degrees of freedom to implement the energy loss. Plus other issues pointed out in \cite{coherbeta, coherbeta2, joydeep} suggest that this may not be a viable option. \label{openqft}}. This means the only way to realize the background \eqref{amhomes} would be as a Glauber-Sudarshan state over a supersymmetric Minkowski background\footnote{Being supersymmetric Minkowski (or more appropriately, warped supersymmetric Minkowski with compact non-K\"ahler internal eight-manifold), there is no longer any Kac-Moody constraints from \cite{sethiK}, or any energy constraints from \cite{GMN}.}. How should we go about constructing such a state now?

Fortunately, since the form of the metric \eqref{amhomes} in M-theory remains unchanged from what we had in \cite{coherbeta, coherbeta2}, the procedure to explicitly construct such a state, will follow \cite{coherbeta2, joydeep}, namely, doing a path-integral over the Minkowski minimum. In other words:

{\footnotesize
\bg\label{wilsonrata5}
\langle {\bf g}_{\mu\nu} \rangle_{\sigma} \equiv {\int [{\cal D} g_{\rm MN}] [{\cal D}{\rm C}_{\rm MNP}] [{\cal D}\overline\Psi_{\rm M}] [{\cal D}
\Psi_{\rm N}]~e^{i{\bf S}_{\rm tot}}~ \mathbb{D}^\dagger({\alpha}, {\beta}, {\gamma}) g_{\mu\nu}(x, y, z)
\mathbb{D}({\alpha}, {\beta}, {\gamma}) \over 
\int [{\cal D} g_{\rm MN}] [{\cal D}{\rm C}_{\rm MNP}] [{\cal D}\overline\Psi_{\rm M}] [{\cal D}
\Psi_{\rm N}]
~e^{i{\bf S}_{\rm tot}} ~\mathbb{D}^\dagger({\alpha}, {\beta}, {\gamma}) 
\mathbb{D}({\alpha}, {\beta}, {\gamma})}, \nd}
where $({\rm M, N, P}) \in {\bf R}^{2, 1} \times {\cal M}_4 \times {\cal M}_2 \times {\mathbb{T}^2\over {\cal G}}$ with ${\cal G}$ being the group action\footnote{The notations used here differ slightly from the ones used earlier after \eqref{morgcome}.}; $x \equiv ({\bf x}, t) \in {\bf R}^{2, 1}, y \in {\cal M}_4 \times {\cal M}_2, z \in {\mathbb{T}^2\over {\cal G}}$; ${\sigma} = ({\alpha}, {\beta}, {\gamma})$ is the Glauber-Sudarshan state associated with 
$(\{{g}_{\rm MN}\}, \{{\rm C}_{\rm MNP}\}, \{{\Psi}_{\rm M}\})$ degrees of freedom (see also 
\cite{coherbeta, coherbeta2}); ${\bf g}_{\mu\nu}$ is the graviton operator related to $g_{\mu\nu}$; the measure ${\cal D} g_{\rm MN} \equiv {\cal D} g_{\mu\nu} {\cal D}g_{\rm AB}$ with $({\rm A, B}) \in {\cal M}_4 \times {\cal M}_2 \times {\mathbb{T}^2\over {\cal G}}$;
$\mathbb{D}(\sigma)$ is a non-unitary displacement operator, {\it i.e.} $\mathbb{D}^\dagger(\sigma) \mathbb{D}(\sigma) \ne 
\mathbb{D}(\sigma) \mathbb{D}^\dagger(\sigma) \ne 1$ 
(see 
\cite{coherbeta} for details);  and the total action:
\bg\label{plazabing}
{\bf S}_{\rm tot} \equiv {\bf S}_{\rm kin} + {\bf S}_{\rm int} + 
{\bf S}_{\rm ghost} + {\bf S}_{\rm gf}, \nd 
where the perturbative part of ${\bf S}_{\rm int}$ comes from an interaction term like eq. (7.55) in \cite{joydeep} and ${\bf S}_{\rm gf}$ is the gauge-fixing term. The ghosts and the gauge-fixing terms are necessary to bring the propagators in the standard forms and remove the redundant degrees of freedom. In writing the expectation value, we have suppressed the measure associated with the ghosts as well as kept the off-shell states massive (while this is not important for the present work, the former will be elaborated in \cite{hete8} and the latter is discussed recently in \cite{joydeep}). 

Unfortunately dealing with the path-integral structure in \eqref{wilsonrata5} with 44 metric degrees of freedom, 84 three-form degrees of freedom and 128 Rarita-Schwinger degrees of freedom (plus the ghost terms) is clearly beyond the scope of the present treatment. We will then follow the simplifying procedure adopted in \cite{coherbeta, coherbeta2, joydeep, borel2}, namely, consider {\it three} scalar degrees of freedom which are the representative samples from the set of 44 gravitons, 84 fluxes and  fermionic condensates from the 128 Rarita-Schwinger fermions. The latter is chosen to avoid Grassmanian integrals in \eqref{wilsonrata5}, and the representative sample of gravitons include the graviton degrees of freedom along the non-compact directions. Even for such simplifying choice, the analysis of the path-integral is still complicated. In \cite{coherbeta, coherbeta2, joydeep, borel2}, it is shown that the path-integral \eqref{wilsonrata5} may be analyzed using the so-called {\it nodal diagrams} instead of the usual Feynman diagrams because of the {\it shifted} vacuum structure. In fact the nodal diagrams show growth of the Gevrey kind \cite{gevrey} implying that Borel resummation \cite{borelborel} of the Gevrey series is now necessary for the path-integral \eqref{wilsonrata5} to make any sense! Putting everything together, and using the computational procedure outlined in \cite{coherbeta, coherbeta2, joydeep, borel2} gives us the following:

{\scriptsize
\bg\label{ruwang}
\langle {\bf g}_{\mu\nu} \rangle_{\sigma} =
\sum_{\{{\bf s}\}}\Bigg[{1\over g_{({\bf s})}^{1/l}}
\int_0^\infty d{\rm S} ~{\rm exp}\left(-{{\rm S}\over g_{({\bf s})}^{1/l}}\right) {1\over 1 - {\cal A}_{({\bf s})}{\rm S}^l}\Bigg]_{\Su {\rm P. V}} 
\int_{k_{\rm IR}}^\mu d^{11}k ~{\overline\alpha_{\mu\nu}(k)\over a(k)}~
{\bf Re}\left(\psi_k({\rm X})~e^{-i(k_0 - \overline\kappa_{\rm IR})t}\right), \nd}
where we restricted the metric function $g_{\mu\nu} = g_{\mu\nu}(x, y)$ with $y \in {\cal M}_4$ only in \eqref{wilsonrata5}; {\Su P. V} is the principal value of the enclosed integral; $g_{({\bf s})}$ is a set of coupling constants defined by inverse powers of ${\rm M}_p$ in a coupling-constant space parametrized by the set ${\{{\bf s}\}}$ of interactions; the first integral is the result of Borel resumming the Gevrey-$l$ growth where $l$ is one less than the total number of fields ({\it i.e.} $l = 2$ here); ${\cal A}_{({\bf s})}$ is the result of computing all the nodal diagrams, including the NLO diagrams for all interactions in the set $\{{\bf s}\}$, that are combinatorially suppressed but {\it not} volume ${\rm V}$ suppressed where the volume ${\rm V}$ refers to the IR volume appearing from the IR/UV mixing \cite{iruv}; $\overline\alpha_{\mu\nu}(k) = {\alpha_{\mu\nu}(k)\over {\rm V}} \equiv {\alpha(k)\eta_{\mu\nu}\over {\rm V}}, a(k) = {k^2\over {\rm V}}$ if we choose the scale to be ${\rm M}_p$ and $\alpha(k)$ is related to the Glauber-Sudarshan state $\vert\sigma\rangle$ discussed earlier; and $\psi_k({\rm X})$, with ${\rm X} \equiv ({\bf x}, y) \in ({\bf R}^2, {\cal M}_4)$, is the spatial wave-function over the solitonic background with $\overline\kappa_{\rm IR} (> k_{\rm IR})$ being an IR scale. (See \cite{coherbeta, coherbeta2, joydeep, hete8, borel2} for details on the aforementioned computations.)

The summing over the set $\{{\bf s}\}$ of interactions in the coupling constant space means that we are summing over Borel-resummed series. This {\it double}-summing is necessary to make the cosmological constant $\Lambda$ {\it small} \cite{joydeep, borel3}, where the closed form expression of the dimensionless cosmological constant\footnote{Recall that both the cosmological constant $\Lambda$ and the conformal time $t$ are made dimensionless here using the scale ${\rm M}_p \equiv 1$. One could use a different scale, namely the size of the internal eight-manifold from the supersymmetric Minkowski background, but the final answer remains unaffected by this choice \cite{borel2}.} appears from the first integral over $d{\rm S}$ in \eqref{ruwang}, namely:
\bg\label{montehan}
{1\over \Lambda^{\kappa}} \equiv \sum_{\{{\bf s}\}}\left[{1\over g_{({\bf s})}^{1/l}}
\int_0^\infty d{\rm S} ~{\rm exp}\Bigg(-{{\rm S}\over g_{({\bf s})}^{1/l}}\Bigg) {1\over 1 - {\cal A}_{({\bf s})}{\rm S}^l}\right]_{\Su {\rm P. V}}, \nd
with all the parameters appearing above being defined earlier, and $\kappa$ will be determined later (see \eqref{palrutarckat}).
This integral form of the cosmological constant has already been shown to be positive definite in \cite{borel2}, irrespective of the signs of ${\cal A}_{({\bf s})}$ and in \cite{joydeep, borel3} we argue that this can be made small. 

All in all, although it may appear that the analysis follows closely the ones from \cite{coherbeta, coherbeta2, joydeep, borel2}, there are few key differences. {\Su First} is the temporal domain of the validity of the Glauber-Sudarshan state, {\it i.e.} the temporal domain where the state remains approximately {\it coherent}, and {\Su second} is the functional form of $\alpha_{\mu\nu}(k)$. Can they be precisely determined here?

The answer turns out to be yes provided we know the functional form for
${\rm F}_1(t)$. This is represented by a series in ${g_s\over {\rm HH}_o}$ in \eqref{sezoe70}. Compared to the type IIB case studied in \cite{desitter2} the situation is different because ${g_s\over {\rm HH}_o}$ itself depends on ${\rm F}_1(t)$ as shown in \eqref{beatrvit}. This means, even before we invoke the consequence from the Schwinger-Dyson's equations, ${\rm F}_1(t)$ has to satisfy:
\bg\label{4mecoffee}
{\rm F}_1(t) = \left(t\sqrt{\Lambda {\rm F}_1(t)}\right)^{\beta_o} 
\sum_{k = 0}^\infty {\rm A}_k \left(t\sqrt{\Lambda {\rm F}_1(t)}\right)^{2k/3}, \nd
where ${\rm A}_k$ are time independent constants; and in the limit ${\rm M}_p \equiv 1$ we take both $\Lambda$ and $t$ to be dimensionless as explained earlier. In addition to this, there is also a derivative constraint coming from the NEC non-violating criterion \cite{coherbeta2} as in \eqref{voluntina}. The coefficients ${\rm A}_k$ can only be fixed using the Schwinger-Dyson's equations (see equivalent example for the type IIB case in \cite{desitter2, coherbeta, coherbeta2, joydeep}). This is in general a non-trivial exercise because of the mixing of the various degrees of freedom (including ghosts), but we can try a toy example. A simple case would be when ${\rm A}_0 >> {\rm A}_k$ for $k \ge 1$. For such a case, defining ${\rm A}_0 \equiv 1$, we find:
\bg\label{julesmeyy}
{\rm F}_1(t) = \left(\Lambda t^2\right)^{\beta_o\over 2 - \beta_o}, ~~~~~ -{1\over \sqrt{\Lambda}} ~ < t ~ \le 0, \nd
with $0 \le \beta_o < 2$ for the system to remain consistent\footnote{Note that, with the choice of ${\rm F}_1(t)$ from \eqref{julesmeyy}, both the 
type IIA coupling ${g_s\over {\rm HH}_o} = \left(\Lambda t^2\right)^{1\over 2 - \beta_o}$ and the heterotic coupling ${g_{\rm het}\over {\rm H}^2} = 
\left(\Lambda t^2\right)^{\beta_o \over 2 - \beta_o}$ are small at late time, {\it i.e.} when $t \to 0$. Thus both the system are naturally {\it weakly} coupled at late time as long as $0 \le \beta_o < 2$. Beyond this regime of $\beta_o$, there are multiple issues with the Glauber-Sudarshan states in both heterotic and the dual type IIB theories.}, and not violate the NEC criterion. Recall that as long as $0 < \beta_o < + {2\over 3}$, the type IIB system can dynamically go to $SO(32)$ heterotic (broken to a suitable subgroup that we will discuss soon). The NEC non-violating criterion allows $0 \le \beta_o < 2$, so is compatible with the aforementioned condition. More so, the size of ${\cal M}_4$ {\it grows} at late time keeping the four-dimensional Newton's constant time-independent in the heterotic side. 

Interestingly the temporal domain of the validity of our analysis matches exactly with the TCC bound advocated in \cite{tcc}, at least for this simple toy example. Additionally, plugging \eqref{julesmeyy} in \eqref{voluntina} $-$ to see whether the non-violating NEC criterion from 
\cite{coherbeta2} is satisfied or not $-$ gives us the following solution:
\bg\label{madinariz}
\gamma_o = {\beta_o\over 2}, ~~~~~ {\rm C}_0 = {2\sqrt{\Lambda}\over 2 - \beta_o}, ~~~~~ {\rm C}_k = 0, ~~~~~ \forall k > 0. \nd
This is clearly consistent as long as $\beta_o \ge 0$. Thus combining the compatibility criteria from both TCC \cite{tcc} and non-violating NEC \cite{coherbeta2}
gives us a stronger reason to justify the existence of a de Sitter space as a Glauber-Sudarshan state in $SO(32)$ heterotic string theory.

Two points still remain. {\Su One}, the functional form for $\alpha_{\mu\nu}(k)$ which would specify the Glauber-Sudarshan state in the supergravity configuration space, and {\Su two}, the form of the vector bundle over the internal six-manifold. The latter is a bit subtle because, while the four-dimensional base of the internal six-manifold in \eqref{kimmetea} is time-independent, the fibration structure {\it is} time-dependent because of its dependence on time-dependent NS two-form fluxes from the dual type IIB side. Nevertheless the vector bundle can be studied from {\it localized} G-flux components whose expectation values may be extracted from the corresponding Glauber-Sudarshan state. The computation should be similar to the path-integral analysis we did for the metric in \eqref{wilsonrata5}, which in turn means Gevrey growth of the corresponding nodal diagrams and the subsequent Borel resummation to extract finite answer. For the simpler case studied here, we can restrict the ${\bf D}_4^4$ bundle, alluded to earlier, on the non-K\"ahler four-dimensional base of the internal six-manifold. Unfortunately a generic study of the vector bundle is beyond the scope of this paper and will be dealt elsewhere\footnote{A generic study has to deal with two aspects of the internal space: one, the temporal dependence of the fibration structure and two, the non-K\"ahler (and subsequently, non-complex) nature of the six-manifold. The latter has been addressed in the past, for example in \cite{DRS}, but the former is new. Other issues like anomaly cancellations and flux quantizations follow the route laid out in \cite{desitter2}. We will discuss more on this in \cite{hete8}.}.

Finally let us figure out the functional form for $\alpha_{\mu\nu}(k)$. It is now related to the Fourier transform of the temporal part of the M-theory metric \eqref{amhomes}. Since we are only dealing with real fields in the path-integral, it suffices to take the cosine Fourier transform. Doing this gives us both the functional form for $\alpha_{\mu\nu}(k)$ and the value of the parameter $\kappa$ in the expression for the cosmological constant $\Lambda$ in \eqref{montehan}. They are:

{\footnotesize
\bg\label{palrutarckat}
\alpha_{\mu\nu}(k) = \sqrt{2\over \pi} \Gamma\Big[1 - {16\over 3(2 - \beta_o)}\Big] ~{\rm sin} \Big[{8\pi\over 3(2 - \beta_o)}\Big]\left[k^{1 + {16\over 3(2 - \beta_o)}} + \chi(k)\right]\eta_{\mu\nu}, ~~~ \kappa = {8\over 3(2 - \beta_o)}, \nd}
where $\chi(k)$ may be determined by restricting $\alpha_{\mu\nu}(k)$ to 
its temporal part. For $\beta_o = 0$ we recover exactly the results of \cite{borel2} showing that the system is consistent. Thus together with 
\eqref{palrutarckat}, \eqref{madinariz}, and \eqref{julesmeyy} we have a 
$SO(32)$ heterotic de Sitter background \eqref{kimmetea} realized as a Glauber-Sudarshan state with a positive cosmological constant given by \eqref{montehan}.


\section{Discussions and conclusions \label{sec3}}

The quest for the existence of a four-dimensional de Sitter space in type II string theories is a non-trivial problem \cite{susha}, partly because of the existence of various no-go theorems, ruling out classical de Sitter vacuum, and partly because of the absence of a well-defined Wilsonian action over an accelerating background, ruling out quantum de Sitter vacuum. (Although arguments can be made in favor of a quantum de Sitter vacuum using open QFTs \cite{vernon}, there are numerous issues with this\footnote{See however recent attempts to realize de Sitter vacuum solution using loop-holes in the no-go theorems \cite{westphal}, or using AdS spaces \cite{mav}. It will be interesting to find some connections to our work.}. See footnote \ref{openqft}.) Similar fate befalls de Sitter vacua in heterotic theories. In this paper we have steered clear of both classical and quantum {\it vacuum} solutions, and instead argued for the existence of de Sitter space as a Glauber-Sudarshan state. Such a state is the closest we can come to a classical solution, yet the analysis is fully quantum. In fact our analysis reveals that the quantum computations are more subtle because of the asymptotic natures of the perturbation series. The Gevrey growths of the perturbation series then allow us to use the powerful machinery of resurgence and Borel resummations to argue for the existence of such a state. 

Our studies have shown that heterotic $SO(32)$ theory does allow such a state to exist within a finite temporal domain given in \eqref{julesmeyy}, which is consistent with the trans-Planckian bound \cite{tcc}. Moreover 
\eqref{madinariz} reveals that the solution is consistent with the non-violating NEC condition of \cite{coherbeta2}. With some mild approximations, one can even achieve a surprisingly precise determination of such a state in \eqref{palrutarckat} with a closed-form expression for the positive cosmological constant in \eqref{montehan}. 

Few detail still remains to be investigated. For example, we haven't been able to analyze the case for the $E_8 \times E_8$ heterotic theory. The duality chain connecting this to M-theory from \cite{horava} certainly looks promising, but the advantage we had from the eight-manifold in M-theory for the present analysis cannot be replicated so easily using seven-manifold for the $E_8$ case. We also haven't said much on moduli stabilization, flux-quantizations, anomaly cancellations or vector bundles either. All of these should follow the path laid out for the type IIB case in \cite{desitter2} because our heterotic background is dual to an orientifold background in type IIB. Nevertheless a direct analysis from the heterotic side is needed. All these and other related issues will be discussed in our upcoming paper \cite{hete8}.

\section*{Acknowledgements:} 

We would like to thank Heliudson Bernardo, Suddhasattwa Brahma, Mir-Mehedi Faruk, Bohdan Kulinich, Evan McDonough, Brent Pym, Mark Van Raamsdonk, Savdeep Sethi and Alexander Westphal for many discussions related to de Sitter space in heterotic string theory.
The work of SA is supported in part by the Simons Foundation  award number 896696.
 The work of KD is supported in part by a Discovery Grant from the Natural Sciences and
Engineering Research Council of Canada (NSERC). The work of AM is supported in part by the Prime Minister’s Research Fellowship provided by the Ministry of Education, Government of India. PR would like to acknowledge the ICTP's Associate programme where progress on the ongoing work continued during her visit as senior associate. For the purpose of open access, the authors have applied a Creative Commons Attribution (CC BY) licence to any Author Accepted Manuscript version arising from this submission.



\begin{thebibliography}{99}

\bibitem{dyson}
F.~J.~Dyson,
``Divergence of perturbation theory in quantum electrodynamics,''
Phys. Rev. \textbf{85}, 631-632 (1952).



\bibitem{unsal}
See M.~Mari\~no,
``Lectures on non-perturbative effects in large $N$ gauge theories, matrix models and strings,''
Fortsch. Phys. \textbf{62}, 455-540 (2014)
[arXiv:1206.6272 [hep-th]], and references therein;
G.~V.~Dunne and M.~\"Unsal,
``Resurgence and Trans-series in Quantum Field Theory: The CP${}^{{\rm N}-1}$ Model,''
JHEP \textbf{11}, 170 (2012), [arXiv:1210.2423 [hep-th]]; 
``Generating nonperturbative physics from perturbation theory,''
Phys. Rev. D \textbf{89}, no.4, 041701 (2014), 
[arXiv:1306.4405 [hep-th]]; 
G.~Basar, G.~V.~Dunne and M.~\"Unsal,
``Resurgence theory, ghost-instantons, and analytic continuation of path integrals,''
JHEP \textbf{10}, 041 (2013), [arXiv:1308.1108 [hep-th]];
S.~Gukov, M.~Mari\~no and P.~Putrov,
``Resurgence in complex Chern-Simons theory,''
[arXiv:1605.07615 [hep-th]];
L.~Di Pietro, M.~Mari\~no, G.~Sberveglieri and M.~Serone,
``Resurgence and 1/N Expansion in Integrable Field Theories,''
JHEP \textbf{10}, 166 (2021)
[arXiv:2108.02647 [hep-th]];
D.~Dorigoni,
``An Introduction to Resurgence, Trans-Series and Alien Calculus,''
Annals Phys. \textbf{409}, 167914 (2019)
[arXiv:1411.3585 [hep-th]].



\bibitem{tcc}
J.~Martin and R.~H.~Brandenberger,
``The Trans-Planckian problem of inflationary cosmology,''
Phys. Rev. D \textbf{63}, 123501 (2001)
[arXiv:hep-th/0005209 [hep-th]];
A.~Bedroya and C.~Vafa,
``Trans-Planckian Censorship and the Swampland,''
JHEP \textbf{09}, 123 (2020)
[arXiv:1909.11063 [hep-th]];
A.~Bedroya, R.~Brandenberger, M.~Loverde and C.~Vafa,
``Trans-Planckian Censorship and Inflationary Cosmology,''
Phys. Rev. D \textbf{101}, no.10, 103502 (2020)
[arXiv:1909.11106 [hep-th]];
S.~Brahma,
``Trans-Planckian censorship conjecture from the swampland distance conjecture,''
Phys. Rev. D \textbf{101}, no.4, 046013 (2020)
[arXiv:1910.12352 [hep-th]].

\bibitem{coherbeta}
S.~Brahma, K.~Dasgupta and R.~Tatar,
``Four-dimensional de Sitter space is a Glauber-Sudarshan state in string theory,''
JHEP {\bf 07}, 114 (2021)
[arXiv:2007.00786 [hep-th]];\\
``de Sitter Space as a Glauber-Sudarshan State,''
JHEP {\bf 02}, 104 (2021)
[arXiv:2007.11611 [hep-th]].




\bibitem{coherbeta2}
H.~Bernardo, S.~Brahma, K.~Dasgupta, M.~M.~Faruk and R.~Tatar,
``Four-Dimensional Null Energy Condition as a Swampland Conjecture,''
Phys. Rev. Lett. \textbf{127}, no.18, 181301 (2021)
[arXiv:2107.06900 [hep-th]];
``de Sitter Space as a Glauber-Sudarshan State: II,''
Fortsch. Phys. \textbf{69}, no.11-12, 2100131 (2021)
[arXiv:2108.08365 [hep-th]].

\bibitem{borel2}
S.~Brahma, K.~Dasgupta, M.~M.~Faruk, B.~Kulinich, V.~Meruliya, B.~Pym and R.~Tatar,
``Resurgence of a de Sitter Glauber-Sudarshan State: Nodal Diagrams and Borel Resummation,''
Fortsch. Phys. \textbf{71}, no.12, 2300136 (2023)
[arXiv:2211.09181 [hep-th]].

\bibitem{joydeep}
J.~Chakravarty and K.~Dasgupta,
``What if string theory has a de Sitter excited state?,''
(To appear in JHEP)
[arXiv:2404.11680 [hep-th]].

\bibitem{borel3}
S.~Brahma, J.~Chakravarty, K.~Dasgupta, F.~ Guo, B.~Kulinich, and A.~Maji,
``Nodal Diagrammar, Borel Resummations and the Smallness of the Positive Cosmological constant" {\it To Appear}. 

\bibitem{gevrey}
M.~Gevrey, ``Sur la nature analytique des solutions des \'equations aux d\'eriv\'ees partielles. Premier m\'emoire", {\it Annales scientifiques de l'\'Ecole Normale Sup\'erieure}, {\bf 35}, 129–190 (1918);
G.~Mittag-Leffler, ``Sur la repr\'esentation arithm\'etique des fonctions analytiques d'une variable complexe", {\it Atti del IV Congresso Internazionale dei Matematici, Roma}; 6–11 (1908).

\bibitem{borelborel}
E.~Borel, ``M\'emoire sur les s\'eries divergentes", Ann. Sci. \'Ec. Norm. Sup\'er., Series 3, {\bf 16}, 9–131 (1899).

\bibitem{hete8}
S.~Brahma, K.~Dasgupta, A.~Maji, B.~Kulinich, P.~Ramadevi and R.~Tatar,
``de Sitter and quasi de Sitter States in ${\rm SO}(32)$ and ${\rm E}_8 \times {\rm E}_8$ Heterotic String Theories,'' {\it To Appear}.



\bibitem{desitter2}
K.~Dasgupta, M.~Emelin, M.~M.~Faruk and R.~Tatar,
``de Sitter Vacua in the String Landscape,''
Nucl. Phys. B \textbf{969}, 115463 (2021)
[arXiv:1908.05288 [hep-th]];
``How a four-dimensional de Sitter solution remains outside the swampland,''
JHEP {\bf 07}, 109 (2021)
[arXiv:1911.02604 [hep-th]];
``de Sitter Vacua in the String landscape: La Petite Version,''
QTS2019
[arXiv:1911.12382 [hep-th]]; K.~Dasgupta, M.~Emelin, E.~McDonough and R.~Tatar,
``Quantum Corrections and the de Sitter Swampland Conjecture,''
JHEP \textbf{01}, 145 (2019) [arXiv:1808.07498 [hep-th]].

\bibitem{dileep}
J.~Chakravarty, K.~Dasgupta, D.~Jain, D.~P.~Jatkar, A.~Maji and R.~Tatar,
``Coherent states in M-theory: A brane scan using the Taub-NUT geometry,''
Phys. Rev. D \textbf{108}, no.8, L081902 (2023)
[arXiv:2308.08613 [hep-th]].


\bibitem{GMN}
G.~W.~Gibbons,
``Aspects of supergravity theories,''
print-85-0061 (Cambridge);
J.~M.~Maldacena and C.~Nunez,
``Supergravity description of field theories on curved manifolds and a no go theorem,''
Int. J. Mod. Phys. A \textbf{16}, 822-855 (2001)
[arXiv:hep-th/0007018 [hep-th]];
G.~W.~Gibbons,
``Thoughts on tachyon cosmology,''
Class. Quant. Grav. \textbf{20}, S321-S346 (2003)
doi:10.1088/0264-9381/20/12/301
[arXiv:hep-th/0301117 [hep-th]];
K.~Dasgupta, R.~Gwyn, E.~McDonough, M.~Mia and R.~Tatar,
``de Sitter Vacua in Type IIB String Theory: Classical Solutions and Quantum Corrections,''
JHEP \textbf{07}, 054 (2014)
[arXiv:1402.5112 [hep-th]];
H.~Bernardo, S.~Brahma and M.~M.~Faruk,
``The inheritance of energy conditions: Revisiting no-go theorems in string compactifications,''
SciPost Phys. \textbf{15}, no.6, 225 (2023)
[arXiv:2208.09341 [hep-th]].

\bibitem{senkd}
A.~Sen,
``F theory and orientifolds,''
Nucl. Phys. B \textbf{475}, 562-578 (1996)
[arXiv:hep-th/9605150 [hep-th]];
K.~Dasgupta and S.~Mukhi,
``F theory at constant coupling,''
Phys. Lett. B \textbf{385}, 125-131 (1996)
[arXiv:hep-th/9606044 [hep-th]].

\bibitem{horava}
P.~Horava and E.~Witten,
``Heterotic and type I string dynamics from eleven-dimensions,''
Nucl. Phys. B \textbf{460}, 506-524 (1996)
[arXiv:hep-th/9510209 [hep-th]]; ``Eleven-dimensional supergravity on a manifold with boundary,''
Nucl. Phys. B \textbf{475}, 94-114 (1996)
[arXiv:hep-th/9603142 [hep-th]].

\bibitem{DRS}
K.~Dasgupta, G.~Rajesh and S.~Sethi,
``M theory, orientifolds and G - flux,''
JHEP \textbf{08}, 023 (1999)
[arXiv:hep-th/9908088 [hep-th]];
K.~Becker and K.~Dasgupta,
``Heterotic strings with torsion,''
JHEP \textbf{11}, 006 (2002)
[arXiv:hep-th/0209077 [hep-th]]; K.~Becker, M.~Becker, K.~Dasgupta and P.~S.~Green,
``Compactifications of heterotic theory on non-K\"ahler complex manifolds. 1.,''
JHEP \textbf{04}, 007 (2003)
[arXiv:hep-th/0301161 [hep-th]]; 
G.~Lopes Cardoso, G.~Curio, G.~Dall'Agata, D.~Lust, P.~Manousselis and G.~Zoupanos,
``Non-K\"ahler string backgrounds and their five torsion classes,''
Nucl. Phys. B \textbf{652}, 5-34 (2003)
[arXiv:hep-th/0211118 [hep-th]]; 
G.~Lopes Cardoso, G.~Curio, G.~Dall'Agata and D.~Lust,
``BPS action and superpotential for heterotic string compactifications with fluxes,''
JHEP \textbf{10}, 004 (2003)
[arXiv:hep-th/0306088 [hep-th]];
K.~Becker, M.~Becker, K.~Dasgupta and S.~Prokushkin,
``Properties of heterotic vacua from superpotentials,''
Nucl. Phys. B \textbf{666}, 144-174 (2003)
[arXiv:hep-th/0304001 [hep-th]]; K.~Becker, M.~Becker, P.~S.~Green, K.~Dasgupta and E.~Sharpe,
``Compactifications of heterotic strings on non-K\"ahler complex manifolds. 2.,''
Nucl. Phys. B \textbf{678}, 19-100 (2004)
[arXiv:hep-th/0310058 [hep-th]]; M.~Becker and K.~Dasgupta,
``K\"ahler versus non-K\"ahler compactifications,''
[arXiv:hep-th/0312221 [hep-th]].

\bibitem{sethiK}
F.~F.~Gautason, D.~Junghans and M.~Zagermann,
``On Cosmological Constants from $\alpha'$-Corrections,''
JHEP \textbf{06}, 029 (2012)
[arXiv:1204.0807 [hep-th]];
D.~Kutasov, T.~Maxfield, I.~Melnikov and S.~Sethi,
``Constraining de Sitter Space in String Theory,''
Phys. Rev. Lett. \textbf{115}, no.7, 071305 (2015)
[arXiv:1504.00056 [hep-th]];



\bibitem{vernon}
R.~P.~Feynman and F.~L.~Vernon, Jr.,
``The Theory of a general quantum system interacting with a linear dissipative system,''
Annals Phys. \textbf{24}, 118-173 (1963);
C.~Agon, V.~Balasubramanian, S.~Kasko and A.~Lawrence,
``Coarse Grained Quantum Dynamics,''
Phys. Rev. D \textbf{98}, no.2, 025019 (2018)
[arXiv:1412.3148 [hep-th]].
S.~Brahma, A.~Berera and J.~Calder\'on-Figueroa,
``Quantum corrections to the primordial tensor spectrum: open EFTs \& Markovian decoupling of UV modes,''
JHEP \textbf{08}, 225 (2022)
[arXiv:2206.05797 [hep-th]].
T.~Colas, J.~Grain and V.~Vennin,
``Benchmarking the cosmological master equations,''
[arXiv:2209.01929 [hep-th]].
C.~P.~Burgess, R.~Holman and G.~Kaplanek,
``Quantum Hotspots: Mean Fields, Open EFTs, Nonlocality and Decoherence Near Black Holes,''
Fortsch. Phys. \textbf{70}, no.4, 2200019 (2022)
[arXiv:2106.10804 [hep-th]];
C.~P.~Burgess, R.~Holman, G.~Kaplanek, J.~Martin and V.~Vennin, ``Minimal decoherence from inflation,''
[arXiv:2211.11046 [hep-th]].



\bibitem{iruv}
A.~G.~Cohen, D.~B.~Kaplan and A.~E.~Nelson,
``Effective field theory, black holes, and the cosmological constant,''
Phys. Rev. Lett. \textbf{82}, 4971-4974 (1999)
[arXiv:hep-th/9803132 [hep-th]]; 
P.~Draper, I.~G.~Garcia and M.~Reece,
``Snowmass White Paper: Implications of Quantum Gravity for Particle Physics,''
[arXiv:2203.07624 [hep-ph]];
T.~W.~Kephart and H.~P\"as,
``UV/IR Mixing, Causal Diamonds and the Electroweak Hierarchy Problem,''
[arXiv:2209.03305 [hep-ph]].

\bibitem{susha}
M.~Cicoli, J.~P.~Conlon, A.~Maharana, S.~Parameswaran, F.~Quevedo and I.~Zavala,
``String Cosmology: from the Early Universe to Today,''
[arXiv:2303.04819 [hep-th]]; U.~H.~Danielsson and T.~Van Riet,
``What if string theory has no de Sitter vacua?,''
Int. J. Mod. Phys. D \textbf{27}, no.12, 1830007 (2018)
[arXiv:1804.01120 [hep-th]].



\bibitem{westphal}
J.~M.~Leedom, N.~Righi and A.~Westphal,
``Heterotic de Sitter beyond modular symmetry,''
JHEP \textbf{02}, 209 (2023)
[arXiv:2212.03876 [hep-th]];
E.~Silverstein,
``Duality, compactification, and $e^{-1/\lambda}$ effects in the heterotic string theory,''
Phys. Lett. B \textbf{396}, 91-96 (1997)
[arXiv:hep-th/9611195 [hep-th]].


\bibitem{mav}
S.~Antonini, P.~Simidzija, B.~Swingle, M.~Van Raamsdonk and C.~Waddell,
``Accelerating cosmology from $\Lambda <0$ gravitational effective field theory,'' JHEP \textbf{05}, 203 (2023)[arXiv:2212.00050 [hep-th]]; Z.~K.~Baykara, D.~Robbins and S.~Sethi,
``Non-Supersymmetric AdS from String Theory,'' SciPost Phys. \textbf{15}, no.6, 224 (2023)[arXiv:2212.02557 [hep-th]].


\end{thebibliography}
\end{document}